\documentclass[aps,prl,twocolumn,superscriptaddress]{revtex4}
\usepackage{graphicx}
\usepackage{color}
\usepackage[tight]{units}
\usepackage{epstopdf}
\usepackage[normalem]{ulem}
\DeclareGraphicsExtensions{.pdf,.jpg,.png,.eps}
\usepackage{amsfonts}
\usepackage{amsmath}
\usepackage{amssymb}
\usepackage{color}
\usepackage{color}
\usepackage{graphicx}
\usepackage{epstopdf}
\usepackage{lineno}
\usepackage{epstopdf}
\usepackage{adjustbox}

\usepackage[T1]{fontenc}
\usepackage{hyperref}
\usepackage{gensymb}

\newcommand{\mn}{$(Mn_{0.2}Co_{0.2}Ni_{0.2}Cu_{0.2}Zn_{0.2})Cr_2O_4$ }
\newcommand{\mg}{$(Mg_{0.2}Co_{0.2}Ni_{0.2}Cu_{0.2}Zn_{0.2})Cr_2O_4$ }

\begin{document}
\title{\textrm{Long-range magnetic ordering and structural phase transition in disordered high-entropy spinel chromites.}}
\author{Sushanta Mandal}
\affiliation{Department of Physics and Materials Science, Thapar Institute of Engineering and Technology, Patiala 147004, India}

\author{Koushik Chakraborty}
\affiliation{UGC-DAE Consortium for Scientific Research, Khandwa Road, Indore 452001, Madhya Pradesh, India}

\author{Isha}
\affiliation{UGC-DAE Consortium for Scientific Research, Khandwa Road, Indore 452001, Madhya Pradesh, India}

\author{Arvind Kumar Yogi}
\affiliation{UGC-DAE Consortium for Scientific Research, Khandwa Road, Indore 452001, Madhya Pradesh, India}

\author{S. D. Kaushik}
\affiliation{UGC-DAE Consortium for Scientific Research Mumbai Centre, R5 Shed, Bhabha Atomic Research Centre, Mumbai, 400085, India}

\author{Sourav Marik}
\affiliation{Department of Physics and Materials Science, Thapar Institute of Engineering and Technology, Patiala 147004, India}

\author{Tirthankar Chakraborty}
\email[]{tirthankar@thapar.edu}
\affiliation{Department of Physics and Materials Science, Thapar Institute of Engineering and Technology, Patiala 147004, India}

\begin{abstract}
\begin{flushleft}

\end{flushleft}

High-entropy spinel oxides provide an excellent platform for investigating entropy-stabilized correlated systems with strong configurational disorder. In this work, we systematically study the temperature evolution of the structural and magnetic properties of Cr-based high-entropy spinels with compositions \mn and \mg. Our results reveal that both systems crystallize in cubic structure with space group \textit{$Fd\overline{3}m$} at room temperature. Each system undergoes antiferromagnetic ordering below the Néel temperatures $ T_N$ = 49 K and 35 K, respectively. Neutron diffraction measurements confirm the emergence of long-range magnetic order with spiral spin arrangement. Both systems exhibit a structural phase transition from cubic \textit{$Fd\overline{3}m$} to orthorhombic \textit{Fddd} symmetry at approximately 55 K and 85 K, respectively. Notably, despite the significant chemical disorder at the A site, both systems undergo transitions analogous to those observed in low entropy spinel systems. This behavior suggests that high configurational entropy may promote global structural stabilization despite local chemical disorder, thereby preserving long-range orderings and the characteristic symmetry-breaking transitions of the pristine spinel systems.

\end{abstract}
\maketitle

\section{Introduction}
Transition metal oxides have received substantial attention in materials science and condensed matter physics due to their novel physical properties, including structural, electrical conductivity, electronic, and magnetism \cite{garg2023co1, shi2025layered, halcrow2013jahn, perveen2025structural, malavekar2024recent, pollini2001electronic}. The complex interplay among spin, lattice, and orbital degrees of freedom governs the emergence of phenomena in such systems \cite{tokura2000orbital, dagotto2005complexity, halcrow2013jahn, khomskii2014transition}. In conventional functional materials, these interactions typically manifest as cooperative Jahn-Teller distortions, multiferroicity, or high-temperature superconductivity. Previously, the discovery of multifunctional systems has been based on the clean lattice paradigm, in which long-range structural or magnetic order is a predictable consequence of stoichiometry and periodic symmetry. 
However, the rise of entropy-stabilized materials has fundamentally shifted this convention \cite{cantor2004microstructural, yeh2004nanostructured, rost2015entropy}. By incorporating five or more primary cations into a single, disordered sublattice, configurational entropy ($\Delta S_{conf}$) is enhanced, which stabilizes phases that are otherwise energetically unfavorable under traditional thermodynamic constraints \cite{rost2015entropy, berardan2016room}. These materials are the so-called high entropy materials. Although high entropy oxides are known for their remarkable thermal stability and strong resistance to phase segregation, they pose a fundamental scientific challenge, such as understanding how long-range symmetry breaking can arise from an environment characterized by extreme local chemical and structural disorder. This issue becomes particularly compelling in the structural family ($AB_2O_4$), where multiple cations occupy crystallographically distinct sites, potentially amplifying the interplay between configurational disorder and cooperative ordering phenomena.  

\begin{figure}
    \centering
    \includegraphics[width=1.1\linewidth]{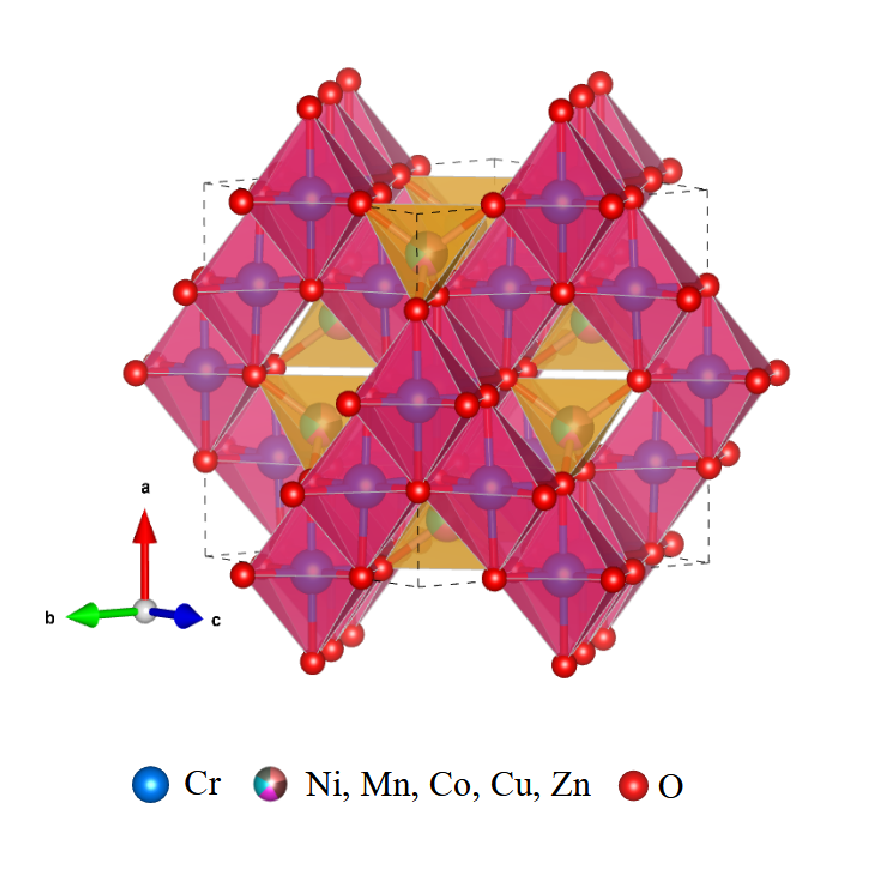}
    \caption{Crystal structure of $(Mn_{0.2}Co_{0.2}Ni_{0.2}Cu_{0.2}Zn_{0.2})Cr_2O_4$. Here, the A-site (Mn, Ni, Co, Cu, Zn) occupies the tetrahedral position, and the B-site (Cr) occupies the octahedral position.}
    \label{fig:enter-label}
\end{figure}

So far, many polycrystalline high entropy oxides with spinel structures have been studied. B. Musico et al. \cite{musico2019tunable} investigated the magnetic ordering of Cr and Fe-based high entropy spinel oxides. The Cr-based compounds exhibited antiferromagnetic (AFM) ordering at low temperatures, whereas the Fe-based samples exhibited ferrimagnetic (FiM) ordering at comparatively high temperatures.  

It has been realized from previous studies that in typical chromite spinels, such as $NiCr_2O_4$ and $CuCr_2O_4$, the orbital degeneracy of tetrahedral A-site cations $Ni^{2+}$ and $Cu^{2+}$ induces cooperative Jahn-Teller ordering, resulting in a tetragonal distortion at high temperatures \cite{suchomel2012spin, ishibashi2007structural, ortega2008low, kemei2013crystal}. Upon cooling, these systems often undergo a structural transition to an orthorhombic phase ($Fddd$ space group), coinciding with the onset of magnetic ordering. This spin-induced symmetry-breaking is well documented in stoichiometric compounds, where it is an indicator of strong magneto-structural coupling.  However, the introduction of high-entropy disorder on the A-site populated by a stochastic mixture of $Mn/Mg, Co, Ni, Cu$, and $Zn$ introduces a level of chemical complexity that should frustrate these long-range transitions. Our previous studies established the successful single-phase synthesis and characterized the fundamental magnetic properties of these complex systems \cite{mandal2025structural}. 
Investigation of the evolution of magnetic and structural properties in such complex systems seems to be interesting. However, such studies are relatively scarce. 
In the present work, we have synthesized high entropy chromite spinels with compositions \mn and \mg, and systematically investigated the evolution of their structural and magnetic properties with temperature. 
Our results reveal that both systems crystallize in a cubic structure with space group \textit{$Fd\overline{3}m$} at room temperature. Each system undergoes antiferromagnetic ordering below the Néel temperatures $ T_N$ = 49 K and 35 K, respectively. Neutron diffraction measurements confirm the emergence of long-range magnetic order with a commensurate spin arrangement, stabilized by lattice distortion. Both systems exhibit a structural phase transition from cubic \textit{$Fd\overline{3}m$} to orthorhombic \textit{Fddd} symmetry at approximately 55 K and 85 K, respectively. This transition is accompanied by the splitting of characteristic diffraction peaks and anisotropic lattice contraction. Both systems exhibit cooperative symmetry-lowering transitions that are comparable to pure spinel systems, despite the chemical instability at the A site. High configurational entropy may allow for global structural stability despite local chemical chaos, retaining long-range order and symmetry-breaking transitions in parent spinel systems. This work shows that high-entropy spinel oxides are a unique family of materials that exhibit the synergistic coexistence of long-range structural periodicity and extreme configurational disorder, rather than being just disordered.

\section{Experimental Details}

 Polycrystalline powder of \mg and \mn were prepared using a conventional solid-state synthesis route. $Co_3O_4$, MnO$_2$, CuO, ZnO, NiO, MgO, and $Cr_2O_3$ (each with a minimum purity of 99.9\%) were mixed in a stoichiometric ratio, and the resultant mixture of powder form was compressed into pellets. These pellets were sintered at 1100 $^\circ$C for 36 hours in a tubular furnace, followed by quenching to room temperature. The sintering process was repeated three times, with intermediate grinding and pelletization to improve homogeneity.

The powder x-ray diffraction patterns of \mn and \mg compounds were collected by using a Rigaku $Cu-K_{\alpha}$  radiation $\lambda$= 1.54182 \AA  diffractometer (Huber make two-circle goniometer) fixed in Bragg-Brentano geometry, with and a secondary beam graphite (002) monochromator and mounted on a $Cu-K_{\alpha}$ rotating-anode x-ray generator which can work at 10 kW. The measurements were performed from $10\degree$ to 70$\degree$ 2$\theta$, with a step width of 0.02$\degree$. The single-phase nature of the synthesized \mn and \mg compounds were confirmed through room temperature (RT) powder x-ray diffraction (P-XRD). Low temperature structural phase transition studies have been done using temperature dependent P-XRD measurements using the same diffractometer, also equipped with closed cycle refrigerator (CCR) based low-temperature setup [cooling down to 11 K]. The single-phase nature of both of the synthesized compounds and the room-temperature lattice-parameters are obtained from the Rietveld refinement using FullProf software \cite{carvajal1990fullprof}, as shown in Figure \ref{fig:lattice parameter variation}. Rietveld refinement reveals cubic phase with space-group \textit{$Fd\overline{3}m$} and the unit-cell parameters for \mn and \mg compounds are found to be a = 8.3261 (3) \AA, a = 8.3261 (3) \AA, and a = 8.3261 (3) \AA and a = 8.3188 (3) \AA, a = 8.3188 (3) \AA, and a = 8.3188 (3) \AA, respectively are consistent with earlier reported data \cite{mandal2025structural}. 

Neutron diffraction (NPD) data for the \mn sample were collected at different temperatures (3 to 35 K) using a multiposition Sensitive Detector (PSD) Focusing Crystal Diffractometer (FCD) provided by the UGC-DAE Consortium for Scientific Research at the Mumbai center, located at the National Facility for Neutron Beam Research (NFNBR) within the Dhruva reactor in Mumbai, India. The measurements were conducted at a wavelength of ($\lambda$) 1.48 \text{\AA}. The vanadium sample holder was exposed directly to the neutron beam. He3-filled linear position-sensitive detectors were used to record diffracted neutrons, utilising the charge division method. . Rietveld refinements of the diffraction patterns (XRD and NPD) were performed using the FullProf suite of software \cite{carvajal1990fullprof}.

Temperature dependent magnetization measurements were performed using a Quantum Design MPMS-3 superconducting quantum interference device (SQUID) under field-cooled (FC) protocol over a temperature range of 4 to 300 K.

\section{Results and Discussion}

\begin{figure*}
    \centering
    \includegraphics[width=1.05\linewidth]{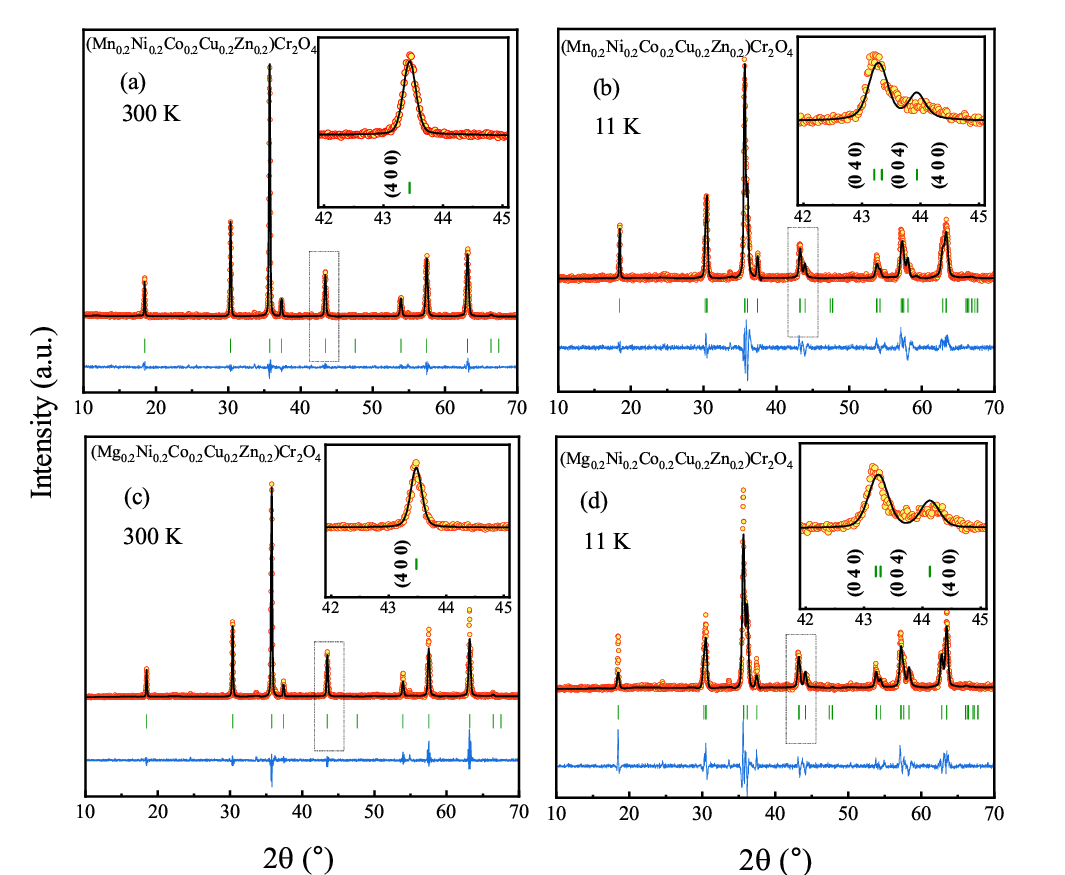}
    \caption{X-ray diffraction patterns of \mn at (a) 300 K and (b) 11 K, and \mg at (c) 300 K and (d) 11 K. Insets in each graph show an enlarged view of the selective peak splitting, where the (4 0 0) peak of the cubic phase splits into (0 4 0), (0 0 4), and (4 0 0) peaks of the orthorhombic phase.}
    \label{fig2-XRD-refinement}
\end{figure*}

X-ray diffraction patterns for both \mn and \mg were collected at various temperatures, ranging from room temperature 300 K down to 11 K. Rietveld refinement was performed on the collected patterns to investigate the structures and their temperature-dependent evolution. During this process, we did not refine the atomic positions and site occupancies, as the weak contribution of oxygen to the XRD patterns could lead to inaccurate results. Therefore, the refinements are effectively equivalent to Le Bail fitting. However, this approach was sufficient to extract the structural parameters and analyze their temperature evolution, which was our primary focus. Representative patterns, along with the refinement at 300 K and 11 K for both, are shown in Figure \ref{fig2-XRD-refinement}. 

At 300 K, the XRD patterns of both systems are well-refined using the cubic space group $Fd\Bar{3}m$. However, at 11 K, the patterns exhibit peak splitting, suggesting a structural transition to a lower symmetry in both systems. In both cases, the structure at 11 K can be refined using the orthorhombic space group $Fddd$. A representative splitting  (4 0 0) of the peak in the cubic phase to (0 4 0), (0 0 4), and (4 0 0) in the orthorhombic phase is shown in the inset of Figure \ref{fig2-XRD-refinement}. 
Details of the refinement results at 300 K and 11 K are provided in Table \ref{Table1-structure-parameters}.  

\begin{table*}
\caption{Structural parameters for \mg and \mn at low temperature and room temperature extracted from Rietveld refinement of powder XRD diffraction data.}
\label{Table1-structure-parameters}
\begin{center}
\small\addtolength{\tabcolsep}{-1pt}

\begin{tabular}{c c c c c} \hline

 & Orthorhombic & Cubic & Orthorhombic & Cubic \\
\hline

 & \mn & & \mg &\\
T  & 11 K & 300 K & 11 K & 300 K \\
Space group  & \textit{Fddd} & \textit{$Fd\overline{3}m$} & \textit{Fddd} & \textit{$Fd\overline{3}m$} \\
a \AA  &  8.2343 (10) & 8.3261 (3) &  8.2017 (8) & 8.3188 (3) \\   
b \AA  &  8.3678 (11) & 8.3261 (3) &  8.3547 (12) & 8.3188 (3) \\
c \AA  &  8.3412 (13) & 8.3261 (3) &  8.3688 (12) & 8.3188 (3) \\
(Mg/Mn/Ni\\/Co/Cu/Zn) &  (0.125, 0.125, 0.125) &   (0.125, 0.125, 0.125) &  (0.125, 0.125, 0.125) &   (0.125, 0.125, 0.125) \\
(x, y, z)\\
B$_{iso}$ & 0.72 (1) & 1.04 (9) & 0.196 (1) & 1.041 (9) \\
Cr (x, y, z) &  (0.5, 0.5, 0.5) &  (0.5, 0.5, 0.5) &  (0.5, 0.5, 0.5) &  (0.5, 0.5, 0.5) \\
B$_{iso}$ & 0.196 (1) & 0.45 (1) & 0.366 (1) & 1.04 (8) \\
O (x, y, z)  &  x = {0.274} (1)  &  x = {0.262} (1) &  x = {0.2521} (9)  &  x = {0.2586} (1) \\
 & y = {0.263} (1) & y = {0.262} (1)  & y = {0.2466} (6) & y = {0.2586} (1) \\
  & z = {0.254} (1) & z = {0.262} (1) & z = {0.2669} (6) & z = {0.2586} (1) \\
$\chi$$^2$ & 2.13 & 1.23 & 2.03 & 1.5 \\
R$_P$ & 14.9 & 8.4 & 19 & 16.7 \\
R$_{WP}$ & 19.4 & 11.8 & 22.9 & 21.9 \\

\\[0.3ex]
\hline\\
\end{tabular}
\par\medskip\footnotesize
\end{center}
\end{table*}

To gain further insight into the structural phase transitions and its onset temperature, we recorded and analyzed XRD patterns at various temperatures between 300 K and 11 K for both systems.

\begin{figure*}
    \centering    
    \includegraphics[width=1\linewidth]{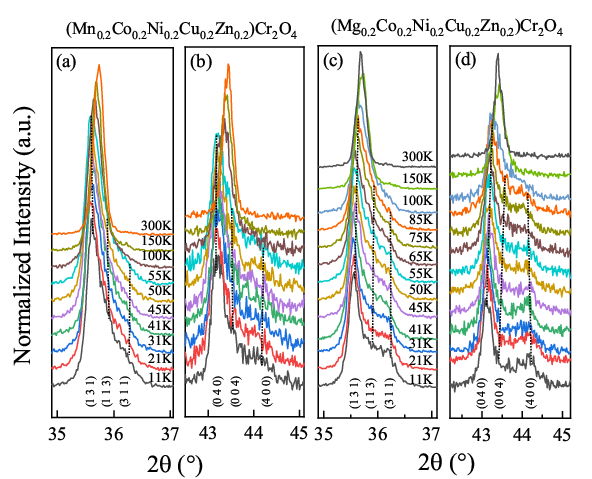}
    \caption{Temperature evolution of XRD patterns for selected peaks of \mn and \mg: (a) The (3 1 1) reflection in the cubic phase splits into (1 1 3), (1 3 1), and (3 1 1) in the orthorhombic phase. (b) The (4 0 0) reflection in the cubic phase splits into (0 4 0), (0 0 4), and (4 0 0) in the orthorhombic phase of \mn. Similar peak splitting for \mg is shown in (c) and (d).}
    \label{fig3-peak evolution}
\end{figure*}

The temperature evolution of a few selective peaks in both systems is shown in Figure \ref{fig3-peak evolution}.

As the temperature is lowered, the peaks corresponding to the cubic phase broaden and begin to split, except for the (1 1 1) reflections. This splitting becomes evident below 55 K for \mn and 100 K for \mg, where the structure is well-described by the orthorhombic space group $Fddd$. In the intermediate temperature range, the cubic phase peaks exhibit significant broadening. We infer that as the temperature decreases, a small fraction of the orthorhombic phase emerges, with its nuclei growing progressively, leading to the development of domain structures and an irregular distribution of domain sizes, and hence to broadening of peaks.
A similar conclusion was drawn for the CuFe$_2$O$_4$ system \cite{balagurov2013structural}. It is noteworthy that in the intermediate temperature range, the XRD data can be refined using a mixed phase of cubic ($Fd\Bar{3}m$) and orthorhombic ($Fddd$).
Nevertheless, the structures above 85 K for \mg and 100 K for \mn can be adequately explained by the cubic space group $Fd\Bar{3}m$ alone, with the orthorhombic phase becoming prominent only below these temperatures. Thus, it can be reasonably concluded that the present systems undergo a cubic-to-orthorhombic phase transition below their respective temperatures. 

\begin{figure*}
    \centering
    \includegraphics[width=1\linewidth]{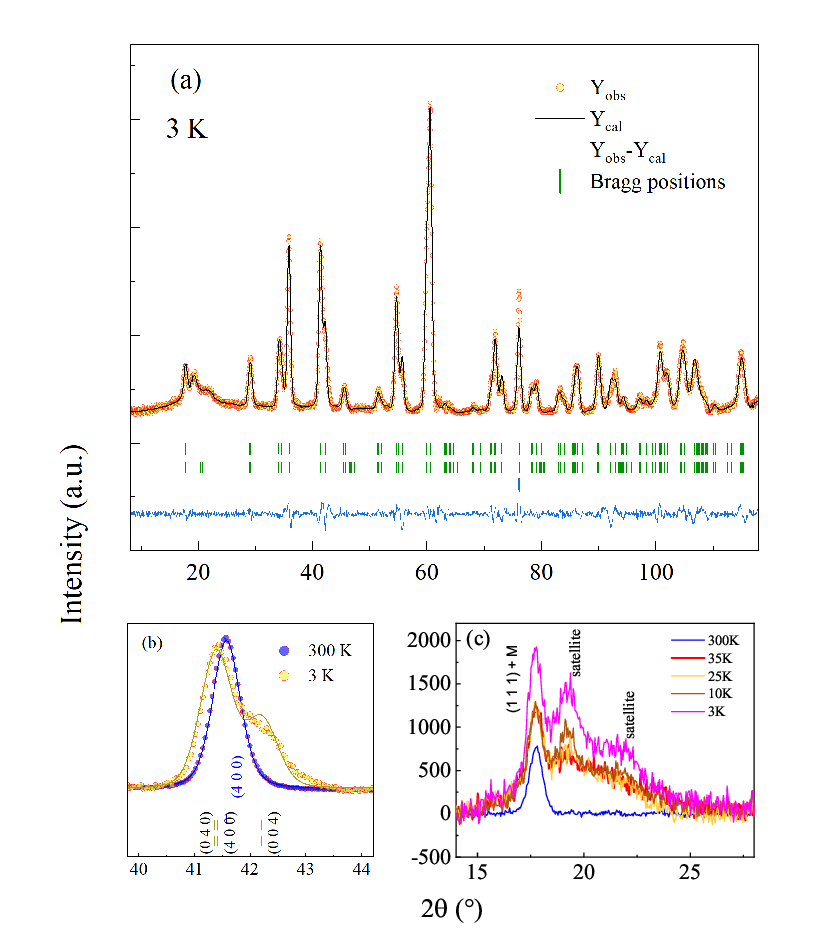}
    \caption{(a) Rietveld refinement of the neutron powder diffraction pattern at 3 K for \mn, showing observed intensity, calculated pattern, difference profile, and Bragg positions corresponding to nuclear and magnetic reflections. (b) show the enlarged (4 0 0) reflection at 300 K as a single cubic peak (blue circle). At 3 K, this peak splits into orthorhombic (0 4 0), (4 0 0), and (0 0 4) reflections, accompanied by the appearance of magnetic Bragg peaks. This confirms magnetically driven symmetry lowering from cubic to orthorhombic. (c) Temperature-dependent neutron diffraction patterns for \mn showing the evolution of magnetic Bragg reflections below the transition ($ T_C$) temperature. The appearance and growth of additional magnetic peaks (including satellite reflections) upon cooling confirm the establishment of long-range antiferromagnetic order. Room-temperature (300 K) NPD results have already been published in ref. \cite{mandal2025structural}}
    \label{fig: NPD_Rietveld_3K}
\end{figure*}

\begin{figure}
    \centering
    \includegraphics[width=1\linewidth]{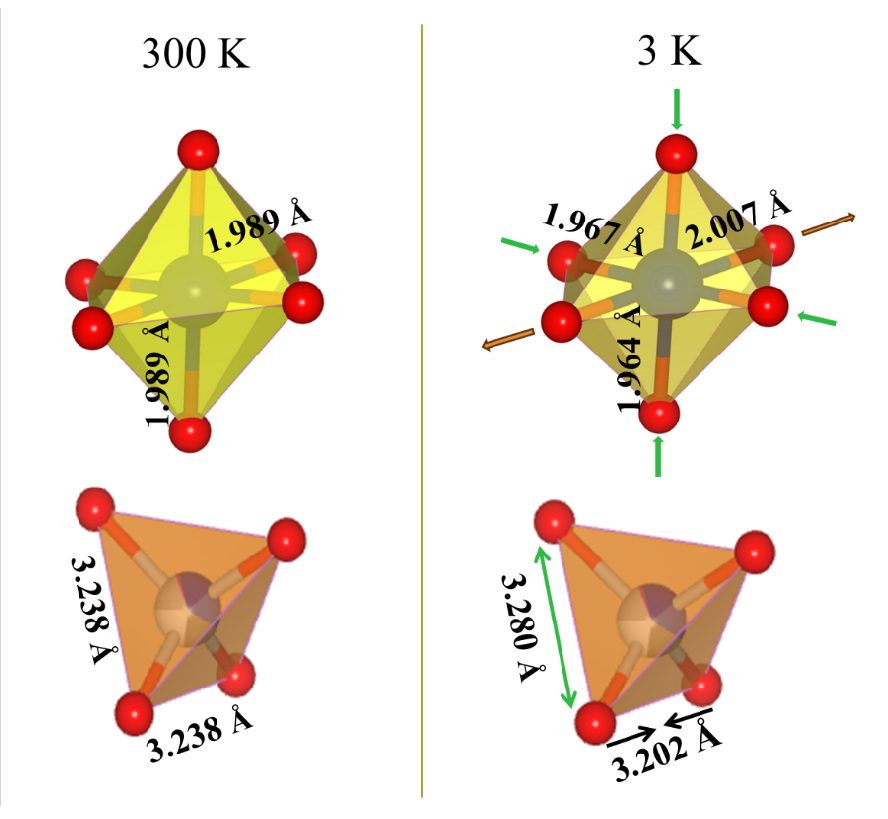}
    \caption{Geometrical representation of the $BO_6$ (B = Cr) octahedral and $AO_4$ (A = Mn, Co, Ni, Cu, and Zn) tetrahedral at temperature 300 K \cite{mandal2025structural} (left side) and 3 K (right side). The arrows indicate the evaluation of bond lengths in the octahedral and tetrahedral at 3 K}
    \label{fig: Tetra_Octa}
\end{figure}

To elucidate the structures and magnetic ground state, neutron diffraction is performed at several temperatures on the \mn system. Figure \ref{fig: NPD_Rietveld_3K} presents the neutron powder diffraction (NPD) results that directly establish the magnetic ground state and its coupling to the structural distortion.
Similar to our XRD refinements, the orthorhombic space group Fddd accurately describes the low-temperature crystal structure of the material. The lattice parameters obtained from the NPD refinements are similar to those obtained using the low-temperature XRD refinements. However, despite the severe disorder and chemical complexity, the 3K NPD data reveal long-range magnetic ordering (Figure 4(c)). Two satellite reflections observed at approximately 19.11 and 21.41 degrees indicate the presence of complex spiral magnetic ordering in the sample. Similar magnetic transitions have also been reported in chromium-based low-entropy spinel oxides \cite{mohanty2020structure}. However, owing to the weak intensity of these peaks, reliable determination of the magnetic structure remains challenging. Consequently, these reflections were excluded from the magnetic refinement and instead treated as background contributions. The refined structural and atomic parameters obtained from the Rietveld refinement of neutron powder diffraction data at 3 K for \mn are summarized in Table \ref{tab:NPD}. Furthermore, such complex non-collinear magnetic ordering may break the spatial symmetry and potentially give rise to ferroelectric behavior. In systems such as $MgCr_2O_4$, $CoCr_2O_4$, $MnCr_2O_4$, and $CuCr_2O_4$, the magnetic transition is often accompanied by a structural distortion and shows incommensurate magnetic structure \cite{mufti2010magnetodielectric, ortega2008low, tomiyasu2004magnetic, mohanty2020structure}. In the present high-entropy system, despite severe A-site chemical disorder, the neutron data clearly demonstrate that long-range periodic magnetic order is preserved. This observation is significant because configurational disorder would typically disrupt coherent spin correlations and favor frustrated states.
The temperature evolution of the magnetic peak intensity follows the typical order-parameter behavior, increasing smoothly with cooling, characteristic of a second-order magnetic phase transition. The persistence of sharp magnetic Bragg peaks down to 3 K confirms that the magnetic coherence length extends over long distances, ruling out cluster-glass behavior in this composition. Also, the long-range magnetic order with strong chemical disorder highlights the remarkable robustness of superexchange-driven magnetism in high-entropy spinel oxides. 

\begin{table}
    \caption{Structural and  atomic parameters obtained from the Rietveld refinement NPD for the \mn system at 3 K.}
    \centering 
    \small\addtolength{\tabcolsep}{-1pt}
    \begin{tabular}{cc}\hline
        Space group & Fddd \\
        a \AA & 8.2207 (6)\\
        b \AA & 8.3708 (6)\\
        c \AA & 8.3818 (2)\\
        Volume \AA$^3$ & 576.7930 (6)\\
        Mn (0.125 0.125 0.125) & \\
        $B_{iso}$ & 0.277 \\
        Occupancy & 0.029 (1) \\
        Ni (0.125, 0.125, 0.125) & \\
        $B_{iso}$ & 0.277 \\
        Occupancy & 0.020 (2) \\
        Co (0.125, 0.125, 0.125) & \\
        $B_{iso}$ & 0.277 \\
        Occupancy & 0.062 (3) \\
        Cu (0.125, 0.125, 0.125) & \\
        $B_{iso}$ & 0.277 \\
        Occupancy & 0.016 (2) \\
        Zn (0.125, 0.125, 0.125) & \\
        $B_{iso}$ & 0.277 \\
        Occupancy & 0.025 (1) \\
        Cr  (0.5, 0.5, 0.5)\\
        $B_{iso}$ & 0.651 \\
        Occupancy & 0.222 (1) \\
        O (x & 0.2658 (2) \\ y & 0.2658 (2) \\z) &  0.2619 (2)\\
        $B_{iso}$ & 0.258 \\
        Occupancy & 0.443 (1)\\
        $\chi^2$ & 9.5 \\
        $R_{P}$ & 4.78 \\
        $R_{WP}$ & 6.48 \\
        \\[0.3ex]
    \hline
    \end{tabular}
    \label{tab:NPD}
\end{table}

Figure \ref{fig: NPD_Rietveld_3K}(b)  shows the representative reflection at 300 K \cite{mandal2025structural} as a single, unsplit cubic (4 0 0) peak, consistent with the cubic (\textit{$Fd\overline{3}m$}) structure. At 3 K, this reflection splits into three distinct peaks indexed as orthorhombic (0 4 0), (4 0 0), and (0 0 4), indicating symmetry lowering to the \textit{Fddd} phase. The observed peak splitting clearly demonstrates a temperature-driven cubic to orthorhombic structural transition.

\begin{figure}[htbp]
    \centering
    \includegraphics[width=1\linewidth]{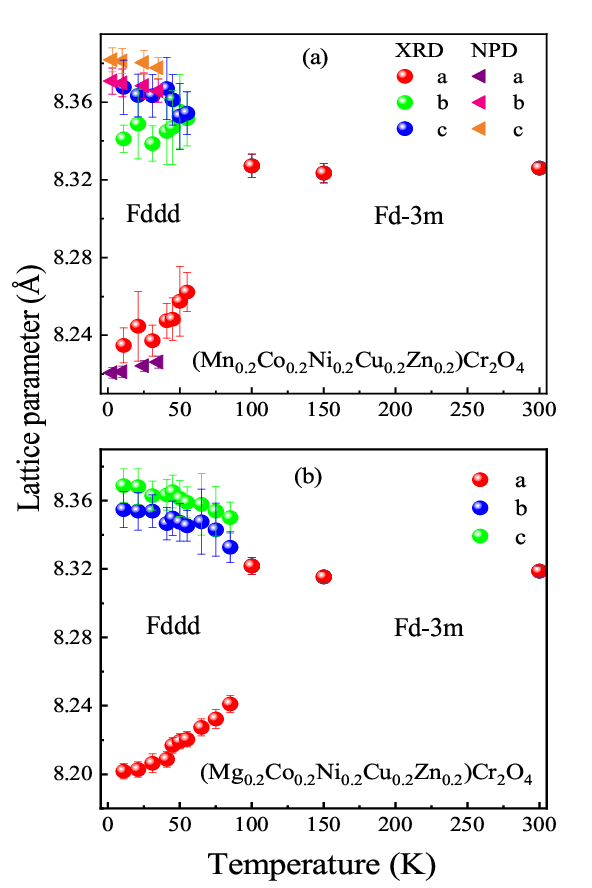}
    \caption{Temperature variation of lattice parameters obtained from Rietveld refinement of (a)\mn on XRD (sphere shape) and NPD (triangle shape)  and (b) \mg on XRD data only.}
    \label{fig:lattice parameter variation}
\end{figure}

The temperature dependence of the lattice parameters extracted from Rietveld refinement of powder XRD for both compositions is presented in Figure \ref{fig:lattice parameter variation}. 
In addition, the lattice constant extracted from Rietveld refinement of neutron powder diffraction data are also represented by triangular symbols for the \mn system. Upon cooling, a clear structural anomaly is observed below approximately 55 K in the \mn system and 85 K in the \mg system, where the high-temperature cubic phase undergoes a symmetry-lowering transition to an orthorhombic structure. This transformation is evidenced by deviations in the lattice parameters from their cubic equivalence and the emergence of distinct orthorhombic splitting, indicating the onset of a low-temperature distorted phase.

It has been observed that in many systems, such as CuCr$_2$O$_4$ and NiCr$_2$O$_4$, structural transitions frequently occur concurrently with the magnetic ordering from the paramagnetic to the ferrimagnetic state. In Ni$Cr_2O_4$, a simultaneous cooperative crystal distortion from tetragonal to orthorhombic symmetry occurs along with the ferrimagnetic transition as reported by Ishibashi and Yasumi \cite{ishibashi2007structural}. Structural changes occur concurrently with magnetic phase transitions in Ni$Cr_2O_4$ and Cu$Cr_2O_4$ are also reported by M. R. Suchomel et al. \cite{suchomel2012spin}. Similar behavior is also reported in the Mg$Cr_2O_4$ system \cite{ortega2008low}. Similarly, the tetragonal to orthorhombic phase transition below $T_N$ in the sample of Mg$Cr_2O_4$ and Zn$Cr_2O_4$ was reported by M. C. Kemei et al. \cite{kemei2013crystal}. This has led to a reasonable conclusion that the magnetic transition is strongly coupled to the lattice degrees of freedom in these systems. In order to check such possibilities in our systems, magnetic susceptibilities as a function of temperature are measured for both systems.

\begin{figure}
    \centering
    \includegraphics[width=1\linewidth]{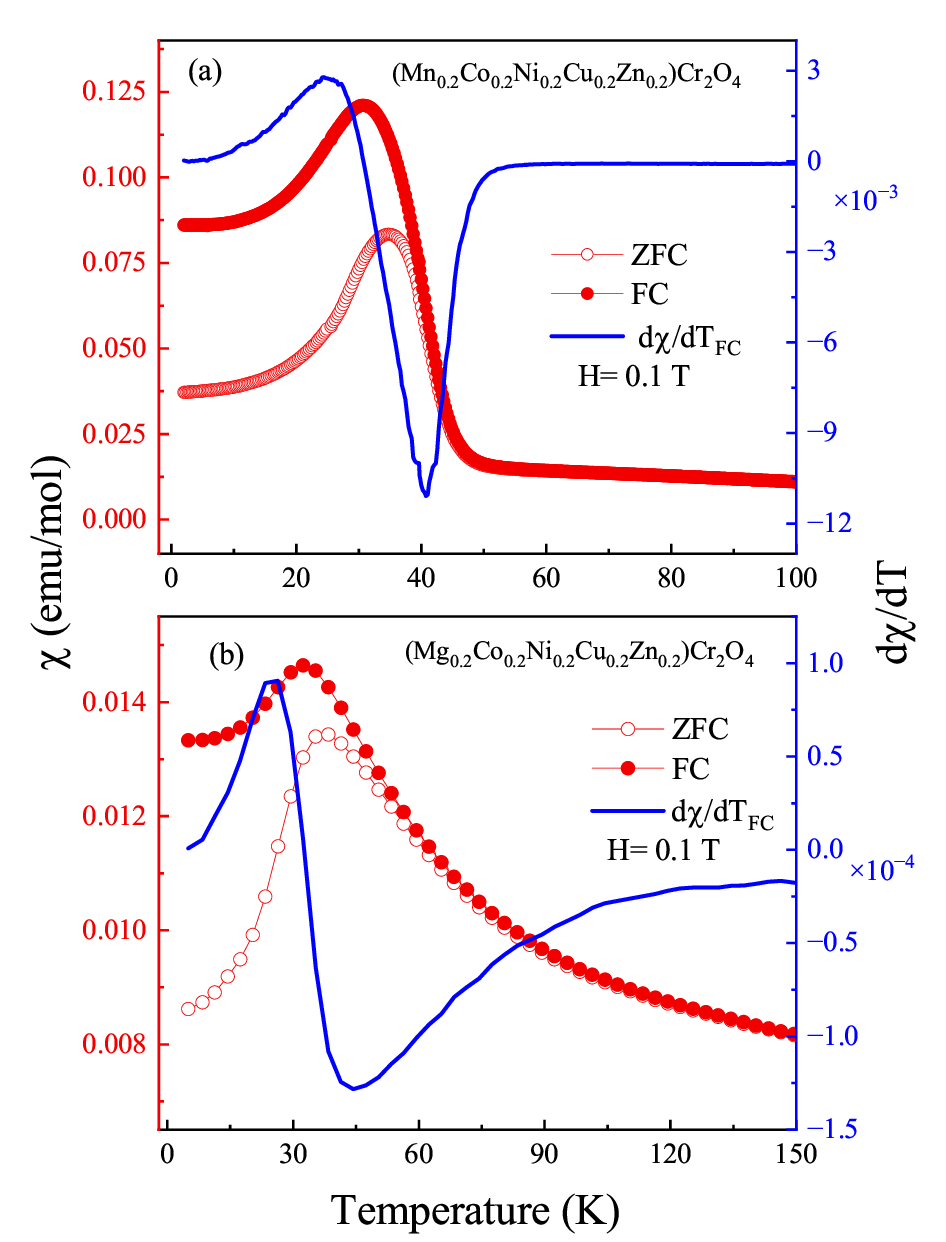}
    \caption{Temperature dependence of dc magnetic susceptibility $(\chi-T)$ measured under an applied field of 0.1 Tesla (T) and corresponding derivative $d\chi/dT$ for (a) \mn and (b) \mg. The sharp anomaly in ($\chi-T$) and peak in $d\chi/dT$ mark the onset of long-range antiferromagnetic ordering.}
    \label{fig:Susceptibility_CR_1_6_0.1T}
\end{figure}
 
Figure \ref{fig:Susceptibility_CR_1_6_0.1T}(left axis) shows the field-cooled (FC) magnetic susceptibility as a function of temperature, measured under an applied magnetic field of  0.1 T over the temperature range 4-100 K. 
The temperature-dependent magnetic susceptibility shows a rapid increase at 49 K for \mn and at 35 K for \mg, corresponding to a paramagnetic to antiferromagnetic transition. The magnetic transition temperatures are determined from the derivative $d\chi/dT$ and are presented in Figure \ref{fig:Susceptibility_CR_1_6_0.1T}(right axis). The sharp peak in $d\chi/dT$ provides estimated of $T_N$, for 49 K and 35 K respectively.
It is worth noting that in systems such as CuCr$_2$O$_4$ and NiCr$_2$O$_4$, which contain Jahn-Teller active cations, a cubic to tetragonal phase transition occurs upon cooling, driven by cooperative Jahn-Teller distortion. With further temperature decrease, these systems undergo a subsequent transition from the tetragonal to the orthorhombic phase. In our case, we did not detect any tetragonal phase in our systems. Whether Jahn-Teller distortion remains cooperative and effective depends on two parameters: temperature and concentration of the Jahn-Teller active cation. In our case, Mn is Jahn-Teller active in the \mn system. However, its concentration is approximately 2.8\%. This may be too small to induce any cooperative Jahn-Teller distortion, which likely explains the absence of a tetragonal phase in our systems. Furthermore, the tetragonal to orthorhombic phase transition in the spinel systems is usually occurs simultaneously and structural and magnetic order parameters are coupled to each other [\cite{suchomel2012spin, ishibashi2007structural}]. However, in both of the present systems, the structural transition temperatures lie well above $T_N$. Usually, in high entropy oxides the short range magnetic interaction persists far above the actual transition temperature. This is a consequence of the inherent disorder present in such system which results in a very gradual change in magnetization with temperature and $1/\chi(T)$ starts to deviate from linearity well above the transition temperature \cite{zhang2019long}. Indeed, such deviation in our systems starts beyond the corresponding phase transition temperatures. Therefore, it may be reasonable to infer that the structural and magnetic order parameters are coupled in these systems also and structural transition takes place where the short range magnetic interaction sets in which might occur even above the actual magnetic transition temperatures. A broad temperature range where the structure can be explained with mixed phases, is thus consistent with this conclusion.

{\color{black}The atomic positions  can be reasonably refined from the neutron diffraction data and the structural changes across the transition can be visulized. Figure \ref{fig: Tetra_Octa} represents the geometrical structure of the constituent polyhedra, namely, the $AO_4$ tetrahedra (where the A-site is disordered and occupied by one of Mn, Co, Ni, Cu, and Zn. In the cubic structure, all the O-A-O bond angles in the $AO_4$ octahedra are the same, ~109.47$\degree$, and all the O-O and A-O bond lengths are also same which form a regular tetrahedron. For the $BO_6$ octahedron, the O-B-O angles deviates significantly from $90\degree$, with one being 93.49$\degree$ and another $86.51\degree$. Such trigonal distortion is closely associated with the oxygen position within the cubic \textit{Fd$\overline{3}m$} space group.} It is worth mentioning here that our results also indicate such trigonal distortion present in the cubic phase of \mn. 
{\color{black} At low temperature, in the $Fddd$ phase, corresponding bond lengths and bond angles changes for both the octahedra and tetrahedra. O-A-O bond angles become 111.083, 110.166, and 107.190 in the $AO_4$ tetrahedra. Also, all of the O-O and A-O bond lengths do not remain the same. On the other hand, $BO_6$ octahedra become more distorted and asymmetric in shape with unequal B-O distances.}

\section{Conclusion}
In our study, we show that the high-entropy chromite spinels \mn and \mg exhibit robust low-temperature magnetostructural phase transitions despite extreme configurational disorder at the A-site. Temperature dependent XRD confirms a clear cubic (\textit{$Fd\overline{3}m$}) to orthorhombic (Fddd) structural transition below 55 K for the Mn-containing system and 85 K for the Mg-containing system, respectively. The absence of an intermediate tetragonal phase suggests that cooperative Jahn-Teller distortion is suppressed due to the dilute concentration of Jahn-Teller-active ions within the entropy-stabilized lattice. Neutron diffraction measurements establish the development of long-range antiferromagnetic ordering below $T_N$ = 49 K and 35 K, evidenced by the emergence of additional magnetic Bragg reflections and their systematic temperature evolution. The preservation of sharp magnetic Bragg peaks down to 3 K implies that long-range magnetic coherence survives strong chemical disorder. 
Thus, high-entropy spinel oxides offer a new platform where disorder and cooperative symmetry breaking coexist.

\section{Acknowledgments}

S. M. acknowledges the SERB, Government of India,
for the SRG/2021/001993 Start-up Research Grant. The authors acknowledge the UGC-DAE Consortium for Scientific Research, Indore, for the Low-temperature XRD. The authors acknowledge the BARC for Neutron Powder Diffraction (NPD).

\bibliography{bib}

@article{yeh2004nanostructured,
  title={Nanostructured high-entropy alloys with multiple principal elements: novel alloy design concepts and outcomes},
  author={Yeh, J-W and Chen, S-K and Lin, S-J and Gan, J-Y and Chin, T-S and Shun, T-T and Tsau, C-H and Chang, S-Y},
  journal={Advanced engineering materials},
  volume={6},
  number={5},
  pages={299--303},
  year={2004},
  publisher={Wiley Online Library}
}

@article{cantor2004microstructural,
  title={Microstructural development in equiatomic multicomponent alloys},
  author={Cantor, Brain and Chang, ITH and Knight, Peter and Vincent, AJB},
  journal={Materials Science and Engineering: A},
  volume={375},
  pages={213--218},
  year={2004},
  publisher={Elsevier}
}

@article{rost2015entropy,
  title={Entropy-stabilized oxides},
  author={Rost, Christina M and Sachet, Edward and Borman, Trent and Moballegh, Ali and Dickey, Elizabeth C and Hou, Dong and Jones, Jacob L and Curtarolo, Stefano and Maria, Jon-Paul},
  journal={Nature communications},
  volume={6},
  number={1},
  pages={8485},
  year={2015},
  publisher={Nature Publishing Group UK London}
}

@article{berardan2016room,
  title={Room temperature lithium superionic conductivity in high entropy oxides},
  author={B{\'e}rardan, D and Franger, S and Meena, AK and Dragoe, N},
  journal={Journal of Materials Chemistry A},
  volume={4},
  number={24},
  pages={9536--9541},
  year={2016},
  publisher={Royal Society of Chemistry}
}

@article{zhang2019long,
  title={Long-range antiferromagnetic order in a rocksalt high entropy oxide},
  author={Zhang, Junjie and Yan, Jiaqiang and Calder, Stuart and Zheng, Qiang and McGuire, Michael A and Abernathy, Douglas L and Ren, Yang and Lapidus, Saul H and Page, Katharine and Zheng, Hong and others},
  journal={Chemistry of Materials},
  volume={31},
  number={10},
  pages={3705--3711},
  year={2019},
  publisher={ACS Publications}
}

@article{musico2019tunable,
  title={Tunable magnetic ordering through cation selection in entropic spinel oxides},
  author={Music{\'o}, Brianna and Wright, Quinton and Ward, T Zac and Grutter, Alexander and Arenholz, Elke and Gilbert, Dustin and Mandrus, David and Keppens, Veerle},
  journal={Physical Review Materials},
  volume={3},
  number={10},
  pages={104416},
  year={2019},
  publisher={APS}
}

@article{kemei2013crystal,
  title={Crystal structures of spin-Jahn--Teller-ordered MgCr2O4 and ZnCr2O4},
  author={Kemei, Moureen C and Barton, Phillip T and Moffitt, Stephanie L and Gaultois, Michael W and Kurzman, Joshua A and Seshadri, Ram and Suchomel, Matthew R and Kim, Young-Il},
  journal={Journal of Physics: Condensed Matter},
  volume={25},
  number={32},
  pages={326001},
  year={2013},
  publisher={IOP Publishing}
}

@article{ortega2008low,
  title={Low temperature neutron diffraction study of MgCr2O4 spinel},
  author={Ortega-San-Martin, L and Williams, AJ and Gordon, CD and Klemme, S and Attfield, JP},
  journal={Journal of Physics: Condensed Matter},
  volume={20},
  number={10},
  pages={104238},
  year={2008},
  publisher={IOP Publishing}
}

@article{suchomel2012spin,
  title={Spin-induced symmetry breaking in orbitally ordered NiCr 2 O 4 and CuCr 2 O 4},
  author={Suchomel, Matthew R and Shoemaker, Daniel P and Ribaud, Lynn and Kemei, Moureen C and Seshadri, Ram},
  journal={Physical Review B},
  volume={86},
  number={5},
  pages={054406},
  year={2012},
  publisher={APS}
}

@article{ishibashi2007structural,
  title={Structural transition of spinel compound NiCr2O4 at ferrimagnetic transition temperature},
  author={Ishibashi, H and Yasumi, T},
  journal={Journal of Magnetism and Magnetic Materials},
  volume={310},
  number={2},
  pages={e610--e612},
  year={2007},
  publisher={Elsevier}
}

@article{tomiyasu2004magnetic,
  title={Magnetic structure of NiCr2O4 studied by neutron scattering and magnetization measurements},
  author={Tomiyasu, Keisuke and Kagomiya, Isao},
  journal={Journal of the Physical Society of Japan},
  volume={73},
  number={9},
  pages={2539--2542},
  year={2004},
  publisher={The Physical Society of Japan}
}

@article{carvajal1990fullprof,
  title={FULLPROF: a program for rietveld refinement and pattern matching analysis, abstracts of the satellite meeting on powder diffraction of the XV Congress of the IUCr},
  author={Carvajal, J},
  journal={Toulouse: France International Union of Crystallography},
  pages={127},
  year={1990}
}

@article{balagurov2013structural,
  title={Structural phase transition in CuFe 2 O 4 spinel},
  author={Balagurov, AM and Bobrikov, IA and Maschenko, MS and Sangaa, D and Simkin, VG},
  journal={Crystallography Reports},
  volume={58},
  pages={710--717},
  year={2013},
  publisher={Springer}
}

@article{pollini2001electronic,
  title={Electronic, spectroscopic and elastic properties of early transition metal compounds},
  author={Pollini, Ivano and Mosser, A and Parlebas, JC},
  journal={Physics Reports},
  volume={355},
  number={1},
  pages={1--72},
  year={2001},
  publisher={Elsevier}
}

@article{malavekar2024recent,
  title={Recent development on transition metal oxides-based core--shell structures for boosted energy density supercapacitors},
  author={Malavekar, Dhanaji and Pujari, Sachin and Jang, Suyoung and Bachankar, Shital and Kim, Jin Hyeok},
  journal={Small},
  volume={20},
  number={31},
  pages={2312179},
  year={2024},
  publisher={Wiley Online Library}
}

@article{perveen2025structural,
  title={Structural, Electronic, and Magnetic Properties of Neutral Borometallic Molecular Wheel Clusters},
  author={Perveen, Saira and Gonzalez Szwacki, Nevill},
  journal={Materials},
  volume={18},
  number={2},
  pages={459},
  year={2025},
  publisher={MDPI}
}

@article{halcrow2013jahn,
  title={Jahn--Teller distortions in transition metal compounds, and their importance in functional molecular and inorganic materials},
  author={Halcrow, Malcolm A},
  journal={Chemical Society Reviews},
  volume={42},
  number={4},
  pages={1784--1795},
  year={2013},
  publisher={Royal Society of Chemistry}
}

@article{shi2025layered,
  title={Layered 3d Transition Metal-Based Oxides for Sodium-Ion and Lithium-Ion Batteries: Differences, Links and Beyond},
  author={Shi, Yuansheng and Hu, Erhai and Sumboja, Afriyanti and Anggraningrum, Ivandini T and Syahrial, Anne Zulfia and Yan, Qingyu},
  journal={Advanced Functional Materials},
  volume={35},
  number={2},
  pages={2413078},
  year={2025},
  publisher={Wiley Online Library}
}

@article{garg2023co1,
  title={The Co1- xZnxCr2O4 [x= 0.0, 0.1] chromite system: A study of structural and frequency dependent room temperature dielectric properties},
  author={Garg, Tarun and Saleem, M and Kaurav, N and Choudhary, P and Yadav, Anand},
  journal={Materials Today: Proceedings},
  volume={89},
  pages={4--7},
  year={2023},
  publisher={Elsevier}
}

@article{mandal2025structural,
  title={Structural, magnetic and x-ray absorption spectroscopy studies of new Cr-based low, medium and high-entropy spinel oxides},
  author={Mandal, Sushanta and Sharma, Jyoti and Chakraborty, Tirthankar and Mahatha, Sanjoy Kr and Marik, Sourav},
  journal={Journal of Alloys and Compounds},
  volume={1010},
  pages={177993},
  year={2025},
  publisher={Elsevier}
}

@article{tokura2000orbital,
  title={Orbital physics in transition-metal oxides},
  author={Tokura, Yoshinori and Nagaosa, Naoto},
  journal={science},
  volume={288},
  number={5465},
  pages={462--468},
  year={2000},
  publisher={American Association for the Advancement of Science}
}

@article{dagotto2005complexity,
  title={Complexity in strongly correlated electronic systems},
  author={Dagotto, Elbio},
  journal={Science},
  volume={309},
  number={5732},
  pages={257--262},
  year={2005},
  publisher={American Association for the Advancement of Science}
}

@article{mufti2010magnetodielectric,
  title={Magnetodielectric coupling in frustrated spin systems: the spinels MCr2O4 (M= Mn, Co and Ni)},
  author={Mufti, N and Nugroho, AA and Blake, GR and Palstra, TTM},
  journal={Journal of physics: Condensed matter},
  volume={22},
  number={7},
  pages={075902},
  year={2010}
}

@article{mohanty2020structure,
  title={Structure and magnetic phase transitions in (Ni1- xCox) Cr2O4 spinel nanoparticles},
  author={Mohanty, P and Venter, AM and Sheppard, CJ and Prinsloo, ARE},
  journal={Journal of Magnetism and Magnetic Materials},
  volume={498},
  pages={166217},
  year={2020},
  publisher={Elsevier}
}

@book{khomskii2014transition,
  title={Transition metal compounds},
  author={Khomskii, Daniel I},
  year={2014},
  publisher={Cambridge University Press}
}

\end{document}